\documentclass[a4,twocolumn]{revtex4}
\usepackage{graphicx}
\usepackage{epstopdf}
\usepackage{longtable}

\begin{document}

\title{Elements of physics for the 21st century}

\author{Werner A. Hofer}
\affiliation{Department of Physics, University of Liverpool, L69 3BX Liverpool, Britain}
\begin{abstract}
Given the experimental precision in condensed matter physics -- positions are measured with errors of less than 0.1pm, energies with about 0.1meV, and temperature levels are below 20mK -- it can be inferred that standard quantum mechanics, with its inherent uncertainties, is a model at the end of its natural lifetime. In this presentation I explore the elements of a future deterministic framework based on the synthesis of wave mechanics and density functional theory at the single-electron level.
\end{abstract}

\maketitle

\section{Introduction}

The paper describes research presented at the EmQM~13 conference. It gives an overview of work on quantum mechanics through about fifteen years, from the first paper on extended electrons and photons published in 1998 \cite{wah1998}, to the last paper on quantum nonlocality and Bell-type experiments in 2012 \cite{wah2012b}. A final section contains the first steps towards a density functional theory of atomic nuclei, presented for the first time at the conference in Vienna. It can be seen that
the publications on quantum mechanics, which I published in this period, possess a gap from about 2002 to 2010. This was due to the realization on my part that I could not account for a simple fact: I could not explain, how the electron changes its wavelength, when it changes its velocity. I felt at the time that not understanding this fact probably meant that I could not understand the electron. Hence I only continued the development of this framework after, prompted by a student of mine, I had found a solution which seemed to make sense. For that I have to thank this particular student. I also have to thank Gerhard Gr\"ossing and Jan Walleczek for organizing this great conference, and the Fetzer Franklin Fund for very generous financial support.

I think we can say today that we actually do understand quantum mechanics. Maybe not in the last details, and maybe not in its full depth, but in the broad workings of the mathematical formalism, the basic physics which it describes, and the deep flaws buried within its seemingly indisputable axioms and theorems. In that, we differ from Richard Feynman, who famously thought that nobody could actually understand it. However, this was said before two of the most important inventions for science in the twentieth century became available to researchers: high-performance computers, and scanning probe microscopes. Computers changed the way science is conducted. Not only do they allow for exquisite experimental control and an extensive numerical analysis of all experiments, they also serve as a predictive tool, if the models include all aspects of a physical system. This, in turn, means that successful theory and successful quantitative predictions, based on local quantities, make it increasingly implausible, that processes exist, which are operating outside space and time. Then, the solution to the often paradoxical theoretical predictions and sometimes incomprehensible experimental outcomes cannot lie in yet another mathematical framework even more remote from everyday experiences than quantum mechanics, but in the rebuilding of a model in microphysics which is both, rooted in space and time, and which allows for a description of single events at the atomic scale. This paper aims at delivering the first building blocks of such a comprehensive model.

\begin{figure}
\includegraphics[width=\columnwidth]{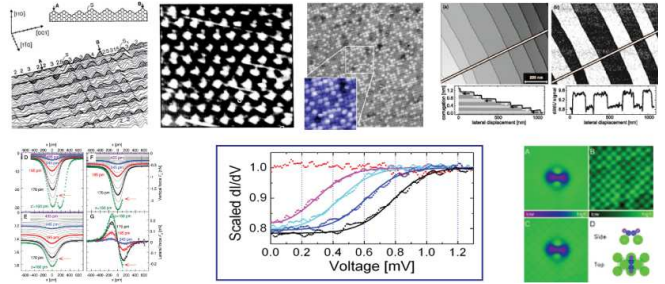}
\caption{Top frames, clockwise, development of scanning probe microscopy over the last 30 years. Au-terraces 1982\cite{binnig1982}, Au-atoms 1987\cite{hallmark1987}, Atomic species 1993\cite{wouda1995}, electron-spin 1998\cite{bode1998}, atomic vibrations 1999\cite{ho1999}, spin-flip excitations 2007\cite{heinrichs2007}, forces on single atoms 2008\cite{ternes2008}.}\label{fig1}
\end{figure}

That we can say today, that we do understand quantum mechanics, is to a large extent the merit of thousands of condensed matter physicists and physical chemists, who have in the last thirty years painstakingly removed layer after layer of complexity from their experimental and theoretical research until they could measure and analyze single processes on individual atoms and even electrons. Today we can measure and simulate the forces necessary to push one atom across a smooth metal surface \cite{ternes2008}, vibrations created by single electrons impinging on molecules \cite{ho1999}, torques on molecules created by ruptures of single molecular bonds \cite{hari2011}, or single spin-flip excitations on individual atoms \cite{heinrichs2007}. See Fig. \ref{fig1} for the development of experiments over the last thirty years.
These experiments have pushed precision to whole new levels. Today, distances can be measured with an accuracy of 0.05 pm \cite{morgenstern2010}, which is about 1/4000th of an atomic diameter, and energies with a precision of 0.1meV, which is about 1/20000th of the energy of a chemical bond \cite{heinrichs2007}. Given these successes and this accuracy, of which physicists could only dream at the time of Einstein, Heisenberg, Schr\"odinger, or Dirac, it would be intellectually deeply unsatisfying if we were today still limited to the somewhat crude theoretical framework of standard quantum mechanics, developed at the beginning of the last century.

The lesson I have learned from my work as condensed matter theorist, trying to make sense of results my experimental colleagues threw at me is this: a physical process has to be thoroughly understood before a suitable theoretical model can be constructed. It is probably one of the more self-defeating features of Physics in the 20th century that new developments mostly took the opposite route: equations came first, processes and physical effects a distant second. This, I think, is about to change again in the 21st century, as mathematical guidance without physical understanding has led Physics thoroughly astray.

\section{The main problem}

The main problem faced by theorists today is the precision of experiments at the atomic scale, because it exceeds by far the limit encoded in the uncertainty relations. This has been the subject of debate for some time now, following the publication of Ozawa's paper in 1988 \cite{ozawa1988}, which demonstrated that the limit can be broken by certain measurements.
An even larger violation can be observed in measurements of scanning probe instruments \cite{wah2012a}. If the instrument measures, via its tunneling current, the variation of the electron density across a surface, then a statistical analysis of such a measurement is straightforward. In the conventional model electrons are assumed to be point particles. The same assumption is made in quantum mechanics, when the formalism is introduced via Hamiltonian mechanics and Poisson brackets. It is also the conventional wisdom in high energy collision experiments, where one finds that the radius of the electron should be less than 10$^{-18}$m. If this is correct, then the density is a statistical quantity derived from the probability of a point-like electron to be found at a certain location. This has two consequences:
\begin{enumerate}
\item A measurement of a certain distance with a certain precision for a particular point on the surface can only be distinguished from the measurement at a neighboring point if the standard deviation is lower than a certain value.
\item A certain energy limit allows only a certain lower limit for the standard deviation in these measurements.
\end{enumerate}
One can now show quite easily \cite{wah2012a} that the standard deviation at realistic energy limits (in case of a silver surface the band energy) is about two orders of magnitude larger than the possible value for state-of-the-art measurements today. The allowed limit for the standard deviation in the experiments is about 3pm, while the standard deviation from the band energy limit is about 300pm. The consequence for the standard framework of quantum mechanics is quite devastating: the uncertainty principle, and by association the whole framework of operator mechanics, becomes untenable, because it is contradicted by experiments. It is precisely this contradiction, which has been claimed by theoretical physicists to be impossible. It also has one consequence, which can be seen as the one principle of the following:
\begin{itemize}
\item The density of electron charge is a real physical quantity.
\end{itemize}
 The density of electron charge has the same ontological status as electromagnetic fields or macroscopic mass or charge distributions. The only difference, and the origin of many of the complications arising in atomic scale physics is that the density not only interacts with external energy sources, but it also interacts with an electron's internal spin density.

The theoretical framework combines two separate models. Both of them are due to physicists born in Vienna, so the location of a workshop on emergent quantum mechanics, from my personal perspective, could not have been better chosen. The first of these physicists is Erwin Schr\"odinger, born in Vienna in 1887, the second one is Walter Kohn, born in Vienna in 1923. The fundamental statements, underlying these two separate models, are the following:
\begin{itemize}
\item A system is fully described by its wavefunction (Schr\"odinger).
\item A system is fully described by its density of electron charge (Kohn).
\end{itemize}
I have been asked, at this workshop, whether the violation of the uncertainty relations could be accounted for by a reduced limit of the constant, e.g. somewhat smaller than $\hbar/2$, a solution which was proposed by Ozawa for the violations detected in the free-mass measurements \cite{ozawa1988}. While this seems, at least for the time being, a possible solution, it disregards the ultimate origin of the uncertainty relations. They are based, conceptually, on the assumption that electrons are point particles (this is the link to classical mechanics and Poisson brackets), and the obligation to account for wave properties of electrons. If wave properties are real, a view taken in the current framework, then there will be no theoretical limit to the precision in their description. A remedy along the lines sketched above then becomes untenable.

\section{Wavefunctions and charge density}
If the density of electron charge is a real physical property, then a common framework must be developed, which allows to map the density onto wavefunctions in the Schr\"odinger theory. Wavefunctions famously do not have physical reality in the conventional model. However, their square does, according to the Born rule. Here, we want to demonstrate that this is correct to some extent also within the new model, but with one important limitation: even though wavefunctions do not have the same reality as mass or spin densities, they can be assembled from these two - physically real - properties.

\subsection{Single electrons}
\subsubsection{Density and energy}
It has been recognized by some of the greatest physicists in the 20th century, among them Albert Einstein, that electrons play a key role in modern physics. Indeed, one could argue that all of physical sciences at the atomic and molecular level, Physics, Chemistry, and Biology is concerned with only one topic: the behavior and properties of electrons. This is also reflected in the celebrated theorem of Walter Kohn: all properties of a physical system, composed of atoms, are defined once the distribution of electron charge within the system is determined \cite{kohn1964}. The solution to the problem of electron density distribution is formulated in density functional theory in a Schr\"odinger-type equation. The spin density is, in this framework, denoted as an isotropic spin-up or spin-down component of the total charge density, the energy related to this spin-density is computed with the help of Pauli-type equations.

However, the framework does not provide physical insights into either spin-densities at the single electron level, or how spin-densities will change in external magnetic fields. What was missing, so far, was a clear connection between the density of electron charge, on the one hand, and the spin density on the other hand. A connection, which should explain the physical origins of wavefunctions in the standard model. It should also explain, how density distributions may change as a consequence of changes to the electron velocity, thus underpinning the wave properties of electrons, found in all experiments.

It turns out to be surprisingly simple to construct such a model. Once it is accepted that electrons must be extended, the wave features must be part of the density distributions of free electrons themselves. In this case the density of charge must also be wavelike. This poses a problem for both, standard quantum theory and density functional theory, because in both cases free electrons are described by plane waves:
\begin{equation}
\psi({\bf r}) = \frac{1}{\sqrt{V}} \exp \left(i {\bf k}{\bf r}\right)
\end{equation}
In this case the Born rule gives a constant value for the probability density, for the mass density and for the charge density: free electrons, then, do not have any distinctive property related to their velocity. But the - now physically real - wave properties of mass and spin densities can be recovered by first assigning a wave-like behavior to the density of electron mass moving in z-direction by:
\begin{equation}
\rho(z,t) = \frac{\rho_0}{2}\left[1 + \cos\left(\frac{4 \pi}{\lambda}z - 4 \pi \nu t\right)\right]
\end{equation}
where $\rho_0$ is the inertial mass density of the electron, and $\lambda$ and $\nu$ depend on the momentum and
frequency according to the de Broglie and Planck rules. At zero frequency infinite wavelength, describing an electron at rest, the mass density is equal to the inertial mass density. However, if the electron moves, then the density is periodic in $z$ and $t$. This requires the existence of an additional energy reservoir
to account for the variation in kinetic energy density. We next introduce the spin density as the geometric product of two field vectors, ${\bf E}$ and ${\bf H}$, which are perpendicular to the direction of motion. These fields are:
\begin{eqnarray}
{\bf E}(z,t) = {\bf e}_1 E_0 \sin\left(\frac{2 \pi}{\lambda}z - 2 \pi \nu t\right) \nonumber \\
{\bf H}(z,t) = {\bf e}_2 H_0 \sin\left(\frac{2 \pi}{\lambda}z - 2 \pi \nu t\right)
\end{eqnarray}
Spin, in this picture, is the geometric product of the two vector components. It is thus a chiral (and for the free electron imaginary) field vector, which is either parallel or anti-parallel to the direction of motion. The total energy density is constant and equal to the inertial energy density if we impose a condition on the spin amplitude \cite{wah2011}:
\begin{eqnarray}
E_{kin}(z,t) &=& \frac{1}{2}\rho_0 v_{el}^2 \cos^2\left(\frac{2 \pi}{\lambda}z - 2 \pi \nu t\right) \nonumber \\
E_{spin}(z,t) &=& \left(\frac{1}{2}\epsilon_0 E_0^2 + \frac{1}{2}\mu_0 H_0^2\right)
\sin^2\left(\frac{2 \pi}{\lambda}z - 2 \pi \nu t\right)  \nonumber \\
&& \left(\frac{1}{2}\epsilon_0 E_0^2 + \frac{1}{2}\mu_0 H_0^2\right) =: \frac{1}{2}\rho_0 v_{el}^2  \\
&\Rightarrow &
E_{tot} = E_{kin}(z,t) + E_{spin}(z,t) = \frac{1}{2} \rho_0 v_{el}^2 \nonumber
\end{eqnarray}
It should be noted that not only the frequency, but also the intensity of the spin component depends on the velocity of the electron. This is in marked contrast to classical electrodynamics, where the energy of a field only depends on the intensity but not on the frequency. Here, it is a necessary consequence of the principle that the electron density is a real physical variable and it establishes a link between the quantum behavior of electrons and the quantum behavior of electromagnetic fields.

This behavior gives a much more precise explanation for the validity of Planck's derivation of black body radiation. If every electromagnetic field, due to emission or absorption of energy by electrons, must follow the same characteristic, then every energy exchange must also be proportional to the frequency of the field. Then Planck's assumption, that $E = h \nu$ is nothing but a statement of this fact. However, that also the intensity follows the same rule, has been unknown so far. In our view this could be the fundamental principle for a general framework of a non-relativistic quantum electrodynamics to be developed in the future.

It should also be noted that the electrostatic repulsion of such an extended electron has to be accounted for, as it is in density functional theory (DFT), by a negative cohesive energy of the electron of -8.16eV. In DFT this energy component is known as the self-interaction correction.

\subsubsection{Wavefunctions}
It is straightforward to assemble wavefunctions from mass and spin density components, following this route. Wavefunctions are in our framework multivectors containing the even elements of geometric algebra in three dimensional space \cite{doran2002}. The even elements are real numbers and bivectors (product of two vectors), the 4$\pi$ symmetry, which is the basis of Fermi statistics in the conventional framework, follows from the symmetry properties of multivectors under rotations in space. The real part $\psi_m$ of a general wavefunction can be written as a scalar part, equal to the square root of the number density:
\begin{equation}
\psi_m = \rho^{1/2} = \rho_0^{1/2} \cos\left(\frac{2 \pi}{\lambda}z - 2 \pi \nu t\right)
\end{equation}
In geometric algebra, this is the scalar component of a general multivector. The bivector component $\psi_s$ is the square root of the spin component, times the unit vector in the direction of electron propagation, times the imaginary unit. It is thus:
\begin{equation}
\psi_s = i {\bf e}_3 S^{1/2} = i {\bf e}_3 S_0^{1/2} \sin\left(\frac{2 \pi}{\lambda}z - 2 \pi \nu t\right)
\end{equation}
The scalar component and the bivector component for an electron are equal to the inertial number density:
\begin{equation}
\rho_0 = S_0 \qquad \Rightarrow \qquad \rho + S = \rho_0 = \mbox{constant}
\end{equation}
The same result can be reached by applying the Born rule, for the wavefunction defined as:
\begin{eqnarray}
\psi &=& \rho^{1/2} + i {\bf e}_3 S^{1/2} \qquad
\psi^{\dagger} = \rho^{1/2} - i {\bf e}_3 S^{1/2} \\
\psi^{\dagger}\psi &=& \rho + S = \rho_0 = \mbox{constant} \nonumber
\end{eqnarray}
The difference to the conventional formulation is that the wavefunction is a multivector, not
a complex scalar. It also makes the spin component a chiral vector, which is important for the understanding of spin measurements.

Formally, we can recover the standard equations of wave mechanics, if we define the Schr\"odinger wavefunction as a complex scalar, retaining the direction of the spin component as a hidden variable.
The wavefunction for a free electron then reads:
\begin{equation}
\psi_S = \rho^{1/2} + i S^{1/2} = \rho_0^{1/2} \exp \left[ i\left(\frac{2 \pi}{\lambda}z - 2 \pi \nu t\right)\right]
\end{equation}
In the conventional framework the dependency of the wavefunction and the Schr\"odinger equation on
external scalar or vector potentials is usually justified with arguments from classical mechanics and
energy conservation. In our approach, the justification is the changed frequency and wavevector of electrons if they are subject to external fields. If we assume that the frequency of the electron varies from the
value inferred from the de Broglie and Planck relations:
\begin{equation}
i \hbar \frac{\partial \psi_S}{\partial t} = h \nu \psi \ne -\frac{\hbar^2}{2m} \nabla^2 \psi_S = \frac{p^2}{2 m} \psi_S,
\end{equation}
then the difference, which is observed in the photoelectric effect, can be accounted for by an additional term in the equation which is linear with the measured scalar potential. Then:
\begin{equation}
i \hbar \frac{\partial \psi_S}{\partial t} = -\frac{\hbar^2}{2m} \nabla^2 \psi_S + V \psi_S
\end{equation}
The second situation, where this can be the case, observed for example in Aharonov-Bohm effects, is when the wavelength does not comply with the wavelength inferred from the frequency and the Planck and de Broglie relations. In this case one can account for the observation by including the vector potential in the differential term of the equation to arrive at the general equation \cite{wah2011}:
\begin{equation}
i \hbar \frac{\partial \psi_S}{\partial t} = \frac{1}{2m} \left(i \hbar \nabla - e {\bf A}\right)^2 \psi_S + V \psi_S
\end{equation}
The important difference, as will be seen presently, is that all these effects occur at a local level and can therefore be analyzed locally: a philosophy, which also forms the core of the local density approximation in DFT.

\subsection{Many-electron systems}
In a many-electron system motion of electrons is correlated throughout the system and mediated by crystal fields within the material. If the spin component in general is a bivector, and if it is subject to interactions with other electrons in the system, then the general, scalar Schr\"odinger equation will not describe the whole physics of the system. Simply accounting for all interactions by a scalar effective potential $v_{eff}$ would recover the Kohn-Sham equations of DFT, if exchange and correlation were included. It would do so, however, for both, density components and spin components, since:
\begin{eqnarray}
\left(-\frac{\hbar^2}{2m} \nabla^2 + v_{eff}\right)\left(\rho^{1/2} + i {\bf e}_3 S^{1/2}\right) = \nonumber \\ = \mu \left(\rho^{1/2} + i {\bf e}_3 S^{1/2}\right) \nonumber \\
\left(-\frac{\hbar^2}{2m} \nabla^2 + v_{eff} - \mu\right)\rho^{1/2} = 0 \\
\left(-\frac{\hbar^2}{2m} \nabla^2 + v_{eff} - \mu\right)S^{1/2} = 0 \nonumber
\end{eqnarray}
In this case the solutions of the equation, single Kohn-Sham states, would exist throughout the system and not lend themselves to a local analysis of physical events. More importantly, such a model would not include an independent spin component in the theoretical description.

We therefore propose a different framework for a many electron system, which scales linearly with the number of electrons and remains local. Such a model can be achieved by including a bivector potential into a generalized Schr\"odinger equation in the following way:
\begin{eqnarray}
\left(-\frac{\hbar^2}{2m} \nabla^2 + v_{eff} + i {\bf e}_v v_b\right)\left(\rho^{1/2} + i {\bf e}_s S^{1/2}\right) \nonumber \\ = \mu \left(\rho^{1/2} + i {\bf e}_s S^{1/2}\right)
\end{eqnarray}
where we have changed the spin component to describe a general spin direction ${\bf e}_s$. The geometric product of two vectors is the sum of a real scalar and an imaginary vector:
\begin{equation}
{\bf e}_v {\bf e}_s = {\bf e}_v \cdot {\bf e}_s - i {\bf e}_v \times {\bf e}_s
\end{equation}
The equation of motion for a general many-electron system then reads:
\begin{eqnarray}
\left(-\frac{\hbar^2}{2m} \nabla^2 + v_{eff} - \mu\right) \rho^{1/2} = {\bf e}_v \cdot {\bf e}_s v_b S^{1/2}
\nonumber \\
\left(-\frac{\hbar^2}{2m} \nabla^2 + v_{eff} - \mu\right) {\bf e}_s S^{1/2} + {\bf e}_v v_b \rho^{1/2}
= \nonumber \\ = - {\bf e}_v \times {\bf e}_s v_b S^{1/2}
\end{eqnarray}
If ${\bf e}_v = 0$, we recover Eq. (13). As inspection shows, the coupled equations only have a solution if the direction of the bivector potential is equal to the direction of spin (${\bf e}_v = {\bf e}_s$), which reduces the problem to:
\begin{eqnarray}
\left(-\frac{\hbar^2}{2m} \nabla^2 + v_{eff} - \mu\right) \left(\rho^{1/2} - S^{1/2}\right) = v_b \left(\rho^{1/2} + S^{1/2}\right) \nonumber \\
\end{eqnarray}
With the transformation $\tilde{\rho}^{1/2} = \rho^{1/2} - S^{1/2}$  and for $v_b = 0$ this equation is identical to the Levy-Perdew-Sahni equation derived for orbital free DFT in the 1980s \cite{lps1986}.
\begin{eqnarray}
\left(-\frac{\hbar^2}{2m} \nabla^2 + v_{eff} - \mu\right) \tilde{\rho}^{1/2} = 0
\end{eqnarray}
One can reduce the expression to the conventional Schr\"odinger equation for the hydrogen atom by
setting $v_{eff} = v_n$, the Coulomb potential of the central nucleus. The equation then has two groundstate solutions, both radially symmetric:
\begin{equation}
\rho^{1/2} = \pm \frac{C}{2} e^{-\alpha r} \qquad S^{1/2} = \pm \frac{C}{2} e^{- \alpha r}
\end{equation}
where $C$ is a constant, and $\alpha$ is the inverse Bohr radius. The vector ${\bf e}_s$ is the
radial unit vector ${\bf e}_r$ and the two spin directions are inward and outward. The same solution
will apply to all $s$-like charge distributions, also, therefore, to the valence electron of silver (see the discussion of Stern-Gerlach experiments below).

The great advantage of the formulation is the simplicity and the reduced number of variables. Both, $\rho$ and $S$ are scalar variables. In addition, we have to find the directions of the unit vectors, ${\bf e}_s = {\bf e}_v$ for every point of the system. This reduces the time independent problem to a problem of finding five scalar components in real space. Compared to the standard formulation of many-body physics, where one has to find a wavefunction of $3N$ variables, where $N$ is the number of electrons, or to standard DFT, which scales cubic with $N$, the approach is much simpler.

However, the effective potential $v_{eff}$ and the bivector potential $v_b$ in this model are generally not known and have to be determined for every system. In standard DFT this is done by calculating the exchange-correlation functional for simpler systems, or for very small systems with high-precision methods. The same route will have to be taken for this new model of many-body physics. Judging from the development of standard DFT this process will probably take at least ten years of development before reliable methods can be routinely used in simulations. But we think, that this method and this approach to many-body physics will also be an element of physics in the 21st century.

\section{Experiments}
As stated in the introduction, we consider the fact that quantum mechanics does not allow for a detailed analysis of single events a major drawback of the theory. However, a theoretically more advanced model will have to pass the test that it can actually deliver these insights. This value statement, i.e. that a theoretical framework is superior not because it obtains higher precision in the numerical predictions, but it is superior because it provides causal insight into physical processes, is somewhat alien to the current debate about quantum mechanics. The tacit agreement seems to be that no theory can provide such an insight. This is one of the fundamental assumptions of the Copenhagen school. There, it is stated {\em that no theoretical model can be more than a coherent framework for obtaining numbers in
experimental trials}. But we do actually not know that this is true, because the assumption that it is true contains an assumption about reality. The assumption that reality cannot {\em in principle} be subjected to an analysis in terms of cause and effect in physical processes. The argument thus is not even logically consistent with its own believe system.

Here, we want to show that the analysis of single events in terms of cause and effect is possible also at the atomic scale. This, we think, demonstrates more than anything else the problems of the standard framework. To an unbiased observer it appears sometimes as if the mathematical tools had, over the last century, acquired a life of their own, so that they are seen as a separate reality, which exists independently of space and time. Hilbert space seems such a concept, and the inability of the standard framework to actually locate a trajectory in space is then seen as proof of the reality of infinitely many trajectories in Hilbert space. This is a logical fallacy, as we shall demonstrate by an analysis of crucial experiments in the following.

\subsection{Acceleration of electrons}
We are quite used to the fact that the wavelength of an electron is inverse proportional to its momentum. It is thus also quite normal to write a wavefunction of a particular free electron, which contains a variation of its amplitude according to this momentum. However, when an electron is accelerated, then standard theory is referring us to the Ehrenfest theorem \cite{ehrenfest1927}. Incidentally, also Paul Ehrenfest was born in Vienna, in 1880. But his theorem only describes the change of an expectation value in a system. It does not allow us to understand, how the wavefunction changes its wavelength, or how the frequency of the wave increases when it interacts with an accelerating potential. Within the present model, this is exactly described at the local level by a new equation, which we call the {\em local} Ehrenfest theorem. Its mathematical expression is:
\begin{equation}
{\bf f} = - \nabla \phi = \rho_0 \frac{d {\bf v}}{dt}
\end{equation}
It states that the force (density) at a particular point of the electron is exactly equal to the gradient of an external potential $\phi$, and that it is described by its classical formulation, the acceleration of its inertial mass. The reason that it is described by this equation is that the number density or the mass density (here we use the two notions interchangeable) is complemented by the spin density to yield a constant:
\begin{equation}
\rho + S = \rho_0 = \mbox{constant}
\end{equation}
The same applies to the square of the wavefunction, which is:
\begin{equation}
\psi^{*} \psi = \rho + S = \rho_0 = \mbox{constant}
\end{equation}
The time differential of momentum density at a particular point is therefore:
\begin{equation}
\frac{d}{dt} \left(m \psi^{*} \psi {\bf v}\right) = m \left(\psi^{*} \psi\right)\frac{d {\bf v}}{dt}
\end{equation}
However, what is hidden in the classical expression is the shift of energy from the mass component to the spin component as the electron accelerates:
\begin{equation}
\dot{S} = - \dot{\rho}
\end{equation}
Here we find the reason for the change of wavelength in an acceleration process: the spin component increases in amplitude, and as gradually more energy is shifted into this component the wavelength becomes shorter and the frequency increases. A process, which so far has remained buried underneath the mathematical formalism and is now open to analysis.

\subsection{Stern-Gerlach experiments}
\begin{figure}
\begin{center}
\includegraphics[width=\columnwidth]{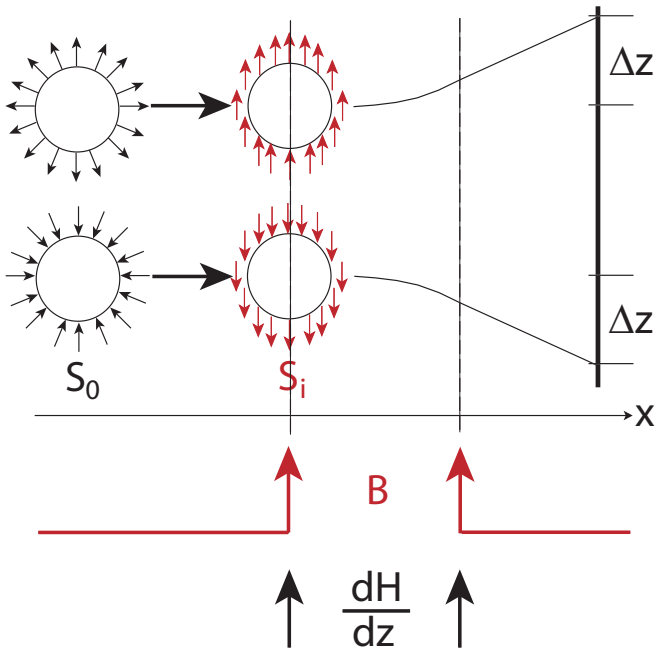}
\caption{Spin measurement of a hydrogen-like atom. Left: the spin densities $S_0$ are parallel to the radial vector. Center: the direction of the induced spin densities $S_i$ is parallel or antiparallel to the magnetic field. Right: due to the inhomogeneous field the atoms are deflected upward or downward.}\label{fig3}
\end{center}
\end{figure}

An inhomogeneous magnetic field leads to deflection of atoms, if they possess a magnetic moment. This effect was used, in the ground breaking experiments on silver by Gerlach and Stern in 1922 \cite{sterngerlach1922}, to demonstrate that the classically expected result, i.e. a statistical distribution around a central impact, is not in line with experimental outcomes. Moreover, the assumption that the orbital moment would cause the deflection was also untenable, because in this case one would observe an odd number of impact locations and not, as in the actual experiments, exactly two. Within the new model the effect is easy to understand. Above, we derived the solution for the electron mass and spin density of a hydrogen-like atom. Assuming that the valence electron of silver can be described in a similar model, we find two different spin directions: one, parallel to the radial vector and directed outward, the other, parallel to the radial vector and directed inward (see Fig. \ref{fig3}, left images).
The induced spin densities $S_i$ (see Fig. \ref{fig3}, centre) as the atoms enter the field, are due to the changes of the spin orientation in a time-dependent magnetic field, which comply with a Landau-Lifshitz like equation \cite{wah2011}:
\begin{equation}
{\bf S} = {\bf e}_S \cdot S \qquad \frac{d {\bf e}_S}{dt} = \mbox{constant} \cdot {\bf e}_S \times
\left({\bf v} \times \frac{d {\bf B}}{dt}\right)
\end{equation}
Then the induced spin densities will lead to a precession around the magnetic field ${\bf B}$ in two directions, which will give rise to induced magnetic moments parallel, or anti parallel to the field. In an inhomogeneous field the force of deflection is then directed either parallel or antiparallel to the field gradient, leading to two deflection spots on the screen, exactly as seen in the experiments (Fig. \ref{fig3}, right). While therefore in the standard model, which assumes that:
\begin{enumerate}
\item Spin is isotropic.
\item A measurement breaks the symmetry of the spin.
\end{enumerate}
no process exists, which could actually explain the symmetry breaking of the initially isotropic spin, the situation is completely different in the new model. Here the process is described by:
\begin{enumerate}
\item Spin is isotropic.
\item The measurement induces spins aligned with the magnetic field.
\item The induced spins lead to positive or negative deflections in a field gradient.
\end{enumerate}
The description is fully deterministic, since the initial direction of spin densities determines the experimental outcome. Statistics only enter the picture, if the initial spin densities are unknown, which they are in practice. Again, we see that the new model actually describes processes at the level of single events, and that probabilities arise due to unknown initial conditions, but that they are not fundamental to a comprehensive model.

\subsection{Interference experiments}
\begin{figure}
\begin{center}
\includegraphics[width=\columnwidth]{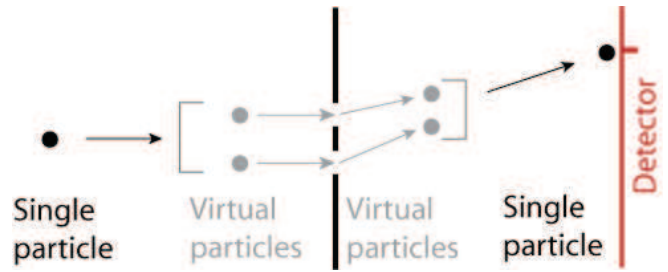}
\caption{Double slit interferometry, Feynman path integrals. A single particle is assumed to split into virtual particles prior to the interferometer. After the interferometer all particles
recombine, the acquired phases along their path determining the interference amplitude. A single particle is detected at the detector screen.}\label{fig4}
\end{center}
\end{figure}

Double slit experiments are notoriously difficult to understand in the framework of standard quantum mechanics. So difficult, in fact, that Richard Feynman called them ''a phenomenon which is impossible, absolutely impossible, to explain in any classical way, and which has in it the heart of quantum mechanics. In reality, it contains the only mystery'' \cite{feynman1964}. The work done recently, aimed at shedding light on this mystery, is already quite convincing: whether it is with Bohm-type trajectories, fluctuating fields \cite{groessing2011}, or whether it is by establishing the trajectories with weak measurements \cite{steinberg2011}, the result always seems to be that one particle passes through one particular opening. Mathematically, the interference phenomena in the standard framework are calculated e.g. with the help of Feynman path integrals. The process described in this mathematical framework is shown in Fig. \ref{fig4}. A single particle, upon entering the vicinity of the interferometer, is assumed to split into a number of virtual particles. Each virtual particle passes exactly one opening of the interferometer, where it acquires a characteristic phase. After the interferometer, all particles are again recombined interfering in a particular way due to their acquired phases. A single impact is observed on the detector screen.

It is quite clear, and is conceded also in the standard framework, that this has nothing to do with real events. However, this insight does not solve the problem, what actually happens so that single entities (electrons or photons), will acquire certain deflections in an interferometer, and why these deflections have an uncanny resemblance to interference patterns of light in an interferometer. In our view, this problem could actually have been solved a long time ago by Duane \cite{duane1916}; a solution which was later taken up by Lande \cite{lande1960}, and which contains no mystery at all.
\begin{figure}
\includegraphics[width=\columnwidth]{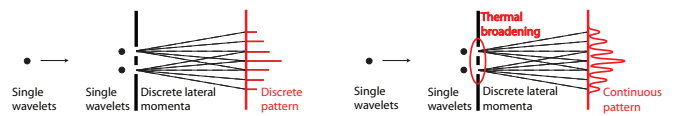}
\caption{Double slit interferometry, real picture. Left: A single particle passing through an opening of the interferometer acquires a discrete lateral momentum due to interactions with the discrete interaction spectrum of atomic scale systems. The interference pattern is a series of sharp impact regions. Right: due to the thermal energy of the slit environment and interactions with molecules in air the impact regions broaden with a Gaussian until they resemble the wave-like interference pattern in an optical interferometer.}\label{fig5}
\end{figure}
The key observation for their model is that every atomic scale system has a discrete interaction spectrum. This means that every interaction of such a system with a single photon or electron can only cause observable changes in the particle's dynamics, if a discrete amount of energy is exchanged, typically corresponding to the excitation of single lattice vibrations. Given this fact, it is impossible that the particle acquires a continuous lateral momentum. Consequently, it also cannot be detected in intermediate regions, unless its trajectory is additionally determined by thermal broadening of the actual interaction.

This model of the process can be experimentally verified. The key to such a verification is the separation of the individual effects changing the particle's trajectory (see Fig. \ref{fig5}). In a liquid helium environment the thermal motion of atoms is frozen. In addition, in an ultrahigh vacuum environment, no interactions with molecules are possible. In a low temperature interferometer the impacts will be sharply defined images of the particle beam deflected by interactions with the atomic environment, while a gradual increase of the temperature of the interferometer should lead to a gradual broadening of the impact regions. This broadening, moreover, should reflect the thermal energy range of the slit environment. Performing such a controlled experiment seems entirely feasible today, and in our view it will establish that indeed the interaction with the atomic environment, and not some fictitious splitting and recombination process is at the bottom of this - hundred years old - mystery.

\subsubsection{Interference of large molecules}
It has been claimed, in a number of high-impact publications since 1999, that large molecules can be made to interfere on gold gratings, and that these experiments show both, the coherence of the molecules over macroscopic trajectories (range of cm), and that the ''wavelength'' of these molecules is equal to the de Broglie wavelength of their inertial mass. This is highly naive and manifestly incorrect, as we show in the following.

As the exemplar of the misguided interpretations we use in the following the first experiments on $C_{60}$ molecules, which were published in the journal {\em Nature} \cite{arndt1999}. Due to the interest of the Chemistry community in these molecules, their properties have been extremely well researched in the past. Theorists routinely calculate their electronic properties, their phonon spectrum, and their light absorption and emission spectrum. They have been adsorbed on surfaces and their charge density distribution has been compared to the results of STM experiments, which verified the theoretical results in great detail. As every Chemist will know, phonon or vibrational modes of organic molecules are varied and range from a few meV (breathing modes, torsion) to a few hundred meV (stretch modes). This particular molecule contains 60 carbon atoms, it thus has 180 modes of vibration which cover the whole energy range.

Experiments are performed in such a way that the molecules are heated with laser light, reaching velocities of a few hundred meters per second, and then passed through a grating with a width of about 50 nm, and a depth of 100 nm. All molecules presented in this type of experiments so far are polarizable, that is they can possess a dipole moment. No control experiments with molecules which are not polarizable have been performed to date. After the grating it is observed that the molecules do not impinge on the screen in a continuous fashion, but that their impact count shows a variation, which is taken as proof that the molecules possess a de Broglie wavelength and interfere as coherent waves.

This is fundamentally wrong on several counts. First, it is well known that the electronic density is fully characterizing a many-electron system. A de Broglie wavelength, which does make sense for free electrons, does not exist in such a structure. Second, it is also well known that internal degrees of freedom of molecular systems start mixing after very short timescales, in the range of femtoseconds. That a molecule is heated with a laser - most likely leading to excitation of electronic transitions - and then spends microseconds preserving a fictitious state vector related to its translational motion, while shaking rapidly due to vibrational excitations is not credible. Third, it is even less credible that such a molecule, with its time dependent dipole moment, will not induce dipole moments in the slit itself, which then interact with the molecule's dipole to alter its trajectory. And fourth, the fictitious state vector of this molecule, which does not exist, is supposed to interfere with another fictitious state vector which went through a different slit, a process, which is completely impossible, unless one assumes that the molecule, during its trajectory, will split into several individual molecules. How this could be possible, given that such a creation of additional molecules violates the energy principle by several MeV, has never been explained and can safely be regarded as pure fiction. In summary, the model is wrong in so many ways, that one is alarmed by the lack of knowledge in basic Chemistry and solid state Physics of its authors and, presumably, the journal's editors.

So how does it really work? Most likely in the way sketched in the previous section. A polarizable molecule is excited by laser light so that most of its low lying vibrational excitations are activated. This molecule enters the interferometer with a time-dependent dipole moment in lateral direction. As the molecule interacts with the atomic environment of the interferometer, it induces electric dipoles into the slit system. These time-dependent dipole moments interact with the molecular dipole moments until the molecule has passed the interferometer. Due to the interaction the molecules acquire a distinct lateral momentum. The momentum leads to a deflection on the detector screen. The deflection is interpreted as the result of a de Broglie wave, because the distance from the point of no deflection to the point of impact is inverse proportional to the velocity of the molecule. Why is it inverse proportional to the velocity of the molecule? Because the time constant of the interaction duration depends on the time the molecule spent in the slit environment of constant depth. Then a faster molecule will spend less time, therefore acquire less lateral momentum, therefore end up closer to the point of no deflection. This, again. has nothing to do with a de Broglie wave, and all to do with the constant distance from the entry to exit of the interferometer (100nm). This whole scenario should be relatively easy to simulate with modern electronic structure methods. One could also try to pin down the actual effect by using non-polarizable molecules. The prediction here is that no periodic variation on the screen will be observed in this case.

\subsection{Aspect-type experiments}
These experiments have been puzzling physicists for at least thirty years. The height of the confusion was probably reached with Aspect's review paper in the journal {\em Nature} in 1999, where he stated: ''The violation of Bell's inequality, with strict relativistic separation between the chosen measurements, means that it is impossible to maintain the image a la Einstein where correlations are explained by common properties determined at the common source and subsequently carried along by each photon. We must conclude that an entangled EPR photon pair is a non-separable object; that is, it is impossible to assign individual local properties (local physical reality) to each photon. In some sense, both photons keep in contact through space and time'' \cite{aspect1999}.

\begin{figure}
\begin{center}
\includegraphics[width=\columnwidth]{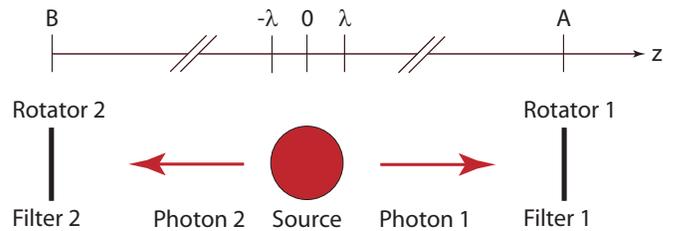}
\caption{Aspect-type experiment. Two photons are emitted from a common source with an initial unknown polarization angle $\varphi_0$. Their polarization is then measured at points $A$ and $B$. (From Ref. \cite{wah2012b}).}\label{fig11}
\end{center}
\end{figure}

We shall show in the following that exactly such a model a la Einstein can explain all experimental data and that the confusion arises from a fundamental technical error in Bell's derivations. To explain the experiments in detail at the single photon level, let us start with setting up a system composed of a source of photons at the point $z = 0$, and two polarization measurements at arbitrary points $z = A$ and $z = -B$. We assume that the polarization measurements contain rotations in the plane parallel to $z$. We also assume that the two photons are emitted from the source with an arbitrary angle of polarization $\varphi_0$. It is irrelevant for the following, whether the field vectors of the two photons rotate during propagation. If they do, this will show up only as an additional angle $\Delta$ between their polarization measurements at $A$ or $-B$. The setup of the experiment is shown in Figure \ref{fig11}.
A single measurement at $A$ consists of two separate processes: First, the polarization angle is altered by an angle $\varphi_A$. Mathematically, this is a rotation in three dimensional space and in the plane perpendicular to the direction of motion, which can be described by the geometric product of a rotator in this plane (a geometric product) $\varphi_A{\bf e}_1{\bf e}_2$ acting on the photon's field vector ${\bf S}$, which is parallel to ${\bf e}_3$. To take care of normalization, we describe such a rotation as:
\begin{equation}
R(A) = \exp \left[\left(\varphi_A +\varphi_0\right){\bf e}_1{\bf e}_2({\bf e}_3)\right] = e^{i\left(\varphi_A + \varphi_0\right)}
\end{equation}
Then, the photon is detected, if the probability $p$ which depends on the angle of rotation and the initial angle of polarization, is larger than a certain threshold:
\begin{equation}
p[R(A)] = \left[\Re (R(A))\right]^2 = \cos^2\left(\varphi_A + \varphi_0\right)
\end{equation}
Depending on how we define our threshold, which is a function of the measurement equipment, an impact at a certain angle of measurement $\varphi_A$ and a certain initial angle $\varphi_0$ is fully determined by the knowledge of these two angles. The single event is thus fully accounted for. However, in the actual experiments the angle $\varphi_0$ is unknown, and it is randomly distributed over the whole interval
[0,$2\pi$]. A set of $N$ experiments will thus lead to a random value for the probability, covering the whole interval [0,1]. The single measurement is thus random. The same is true for a measurement at point $-B$. Also here the polarization measurement is described by a rotation, with a different and fully independent angle $\varphi_B$. The probability of detection is, along the same lines:
\begin{equation}
p[R(B)] = \left[\Re (R(B))\right]^2 = \cos^2\left(\varphi_B + \varphi_0\right)
\end{equation}
Also in this case the single event is fully accounted for if the initial angle $\varphi_0$ and the angle of polarization $\varphi_B$ are known. Again, a set of $N$ experiments will lead to a random value for the probability, covering the whole interval [0,1].

Naively, one could now assume that the correlation probability is the product of the two measurement probabilities at points $A$ and $-B$, respectively. This is exactly what Bell assumed in the derivation of his inequalities, when he wrote \cite{bell1964}:
\begin{equation}
P({\bf a},{\bf b}) = \int d\lambda \rho(\lambda) A({\bf a},\lambda) B({\bf b},\lambda)
\end{equation}
Here, $\lambda$ has the same meaning as the initial angle $\varphi_0$, and the crucial error lies in the assumption that the correlation probability is the product of individual probabilities. This is manifestly incorrect, because it disregards the mathematical properties of rotations. Two separate rotations at $A$ and $-B$ have to be accounted for by a product of individual rotations, thus:
\begin{eqnarray}
R(A) \cdot R(B) &=& \exp \left[\left(\varphi_A +\varphi_0\right){\bf e}_1{\bf e}_2({\bf e}_3)\right] \nonumber \\
&\cdot& \exp \left[\left(- \varphi_B - \varphi_0\right){\bf e}_1{\bf e}_2({\bf e}_3)\right] \nonumber \\
&=& \exp \left[i\left(\varphi_A - \varphi_B\right)\right]
\end{eqnarray}
It is impossible, from these two rotations, to derive a probability which is the product of two positive numbers. Furthermore, the hidden variable $\varphi_0$, which is present in the probability of individual polarization measurements, is canceled out in the correlation derived from two separate rotations. The correct form of the probability for the correlation derived from the two rotations will be:
\begin{equation}
p[R(A),R(B)] = \left[\Re \left(R(A)\cdot R(B)\right)\right]^2 = \cos^2\left(\varphi_A - \varphi_B\right)
\end{equation}
These probabilities are equal to the correlation probabilities derived in the Clauser-Horne-Shimony-Holt formalism \cite{chsh1969}:
\begin{eqnarray}
C^{++} &=& C^{--} = \cos^2\left(\varphi_A - \varphi_B\right) \nonumber \\
C^{+-} &=& C^{-+} = 1 - \cos^2\left(\varphi_A - \varphi_B\right)
\end{eqnarray}
They lead to the standard expectation values measured in Aspect-type experiments:
\begin{equation}
E\left(\varphi_A,\varphi_B\right) = \cos\left[2\left(\varphi_A - \varphi_B\right)\right]
\end{equation}
And they violate the Bell inequalities in the exact same way as found in the experiments:
\begin{eqnarray}
S\left(\varphi_A,\varphi_A',\varphi_B,\varphi_B'\right) &=& E(\varphi_A,\varphi_B) - E(\varphi_A,\varphi_B') + \\ &+& E(\varphi_A',\varphi_B) + E(\varphi_A',\varphi_B') = 2 \sqrt{2} \nonumber
\end{eqnarray}
if  $\varphi_A = 0, \varphi_A' = 45, \varphi_B  = 22.5, \varphi_B' = 67.5$. To repeat the findings: a model based on polarizations and rotations in space recovers all experimental results. It allows for a cause-effect description of every single measurement. It also violates the Bell inequalities. Not, because it is a non-local model, but because Bell made a fundamental error in the derivation of his inequalities. It is thus, paraphrasing Aspect's words, not a proof that a model a la Einstein is impossible, but rather a proof that many quantum theorists do not understand geometry.

\section{Towards a density model of atomic nuclei}
\subsection{Electrons and neutrons}
\begin{figure}
\begin{center}
\includegraphics[width=\columnwidth]{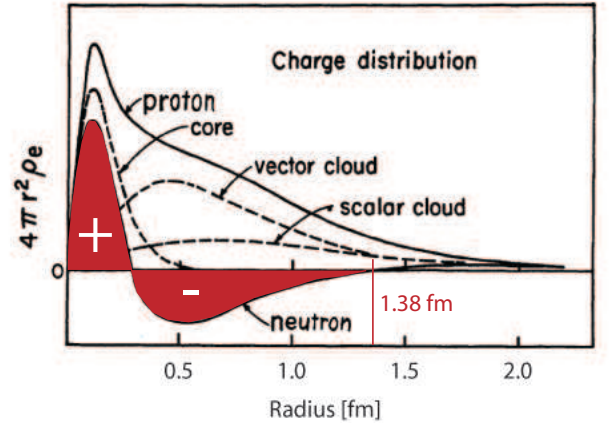}
\caption{Neutron scattering experiments by Littauer et al. \cite{littauer1961}. A neutron consists of a positive core and a negative shell. The radius of the neutron is about 1.38fm. }\label{fig12}
\end{center}
\end{figure}

At the end of this presentation I would like to report on some work in progress. It is quite natural, if one considers the electron an extended particle, to ask, what shape and form it might have apart from the atomic environment. We know from DFT that its density, consequently its shape, will depend on the potential environment. After all, we find much higher densities of electron charge in heavier atoms with a higher number of central charges than we find in hydrogen. So one may also ask, in what shape and form an electron exists, for example, in a neutron. We know that the neutron decays outside an atomic nucleus in about 880 seconds to a proton and an electron, with an excess energy of 785 keV, which is mostly converted into X-ray radiation.
\begin{equation}
n^{0} \rightarrow p^{+} + e^{-} + 785\mbox{keV}
\end{equation}
We also know, from scattering experiments (see Fig. \ref{fig12}), that a neutron contains a core of positive charge, which one could tentatively call the ''proton'' and a shell of negative charge, which one could to first instance identify as the ''electron''.
If the electron exists in such a high density phase, then one could also seek its eigenstates with the help of a Schr\"odinger equation adapted to the much smaller lengthscales and much higher energy scales. However, for such an assumption to make sense it first has to be determined, where the additional mass of the neutron compared to isolated protons and electrons comes from. Here, it has to be remembered that the radius of a neutron is much smaller than the radius of a hydrogen atom. Therefore, the electrostatic field of an electron outside hydrogen has a very low energy of about 11 eV, while this field has a large energy of close to 1 MeV for an electron with a radius of 1.38 fm:
\begin{eqnarray}
W_0^{e} &=& \frac{1}{2} \int_{\infty}^{r_e} \epsilon_0|{\bf E}|^2 dV = \frac{1}{4 \pi \epsilon_0} \frac{e^2}{r_e} \approx 11 \mbox{eV} \nonumber \\
W_n^{e} &=& \frac{1}{4 \pi \epsilon_0} \frac{e^2}{r_n} \approx 1040 \mbox{keV}
\end{eqnarray}
 Here, one finds that the electrostatic energy alone, considering mass equivalents, can account for the excess mass. Next, it is necessary to analyze nuclear units. We know from atomic physics that atomic units are defined from fundamental constants and determine the solution of the hydrogen problem with the Schr\"odinger equation. Let me just remind the reader that an exponentially decaying wavefunction $\psi(r) = \rho_0^{1/2} \exp(-\alpha r)$ leads to the following characteristic equation and the solution for $\alpha$:
\begin{eqnarray}
\left( - \frac{\hbar^2 \alpha^2}{2m} + \frac{2 \hbar^2 \alpha}{2m r} - \frac{e^2}{4 \pi \epsilon_0 r}\right) \psi(r) = E \psi(r) \nonumber \\
\frac{2 \hbar^2 \alpha}{2m r} - \frac{e^2}{4 \pi \epsilon_0 r} = 0 \quad \rightarrow \quad \alpha = \frac{m e^2}{4 \pi \epsilon_0 \hbar^2}
\end{eqnarray}
If a similar solution exists for the neutron, then the decay constant must be different. We account for this hypothesis by rescaling the Planck constant in a nuclear environment so that:
\begin{equation}
\hbar_n = x \hbar \qquad \alpha_n = \frac{1.89 \times 10^{-10} m^{-1}}{x^2} \qquad E_n = \frac{E_H}{x^2}
\end{equation}
The Schr\"odinger equation in a nuclear environment then reads:
\begin{equation}
\left(- \frac{1}{2} \nabla^2 - \frac{1}{r}\right) \psi_n(r) = E_n \psi_n(r)
\end{equation}
The total energy is the sum of the positive energy of the electrostatic field and the negative energy of the eigenvalue, it is known to be 785 keV. It depends, ultimately, on only two values: the radius of the neutron, which is known from scattering experiments, and the scale $x$. With $a_0$ the Bohr radius we get:
\begin{equation}
W_n = \frac{e^2}{4 \pi \epsilon_0 a_0}\left(\frac{a_0}{r_n} - \frac{1}{2 x^2}\right) =
\left(\frac{a_0}{r_n} - \frac{1}{2 x^2}\right) \times 27.211 \mbox[eV]
\end{equation}
The scale $x$ can therefore be calculated from experimental values. With $r_n = 1.38$ fm and $W_n = 785$ keV we get for the scale $x$ and the energy scale $E_n$:
\begin{equation}
x = \frac{1}{18779^{1/2}} = \alpha_f \qquad E_n = 511 \mbox{keV} = m_e c^2
\end{equation}
Both of these values are very fundamental. In the standard model the fine structure constant $\alpha_f$ describes the difference in coupling between nuclear forces and electrostatic forces, while the rest energy of the electron $E_n$ is one of the fundamental constants in high energy physics. At present, we do not have a clear indication of the significance of this finding. It is quite improbable, that this result should be a mere coincidence. After all, the identity relies on two experimental values, the radius of the neutron and the mass of the neutron. Had these values been different, the fine structure constant or the rest energy of the electron would not have been the result of this derivation. We expect that a nuclear model on the basis of high-density electrons, which we also tentatively assume to be an element of physics in the 21st century, will be able to answer this important question.

\subsection{Magic nuclei}
\begin{figure}
\includegraphics[width=\columnwidth]{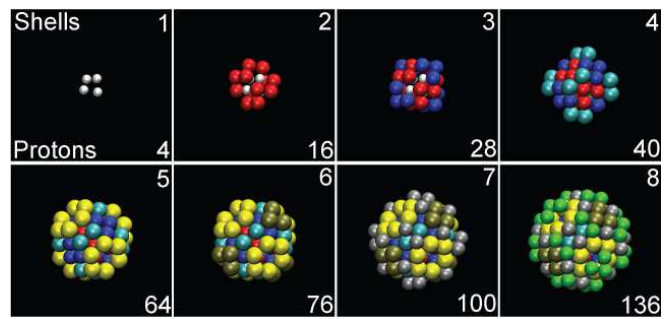}
\caption{Closed shells of atomic nuclei for up to 136 protons. The shell model is only based on geometry and does not include detailed interactions at this point.}\label{fig13}
\end{figure}

It is known that certain numbers of nucleons, assumed to be protons and neutrons in the conventional model, lead to increased stability of atomic nuclei. If high-density electrons are the glue that holds protons together, then protons in a nucleus will be in a regular arrangement. In this case the problem of nuclear organization becomes to first instance a problem of three dimensional geometry. Starting from a single proton, and adding one proton after the other, always under the condition that the distances between protons are constant, will automatically lead to a shell model of atomic nuclei, where a certain number of protons corresponds to closed shells. In Fig. \ref{fig13} we show the first seven closed shells. In particular the first four, with 4, 16, 28, and 40 protons, correspond to magic nuclei in nuclear physics. Larger shells do not necessarily, but it has to be considered that we do not yet have a comprehensive model of interactions within an atomic nucleus, which could account for the observed nuclear masses. Compared to DFT the additional complication within a nucleus is the relatively large volume of protons, which probably cannot be taken into account with a model of point charges, and the unknown role of nuclear forces. Also, it is quite unclear at present if the electrostatic interactions within the nucleus have the same intensity as in a vacuum, how screening works, and what role the energy of electrostatic fields will play in the overall picture. The first steps towards such a model are therefore highly tentative and it is to be expected that a fully quantitative model of atomic nuclei is still a long time in the future. However, such a model could provide a unified basis for discussions in nuclear physics, which connects it seamlessly to other fields of Physics: something, which is manifestly not the case at present.

\section{Summary}
In this presentation I have emphasized six results obtained within a theoretical framework which seamlessly combines wave mechanics and density functional theory. These six results are:
\begin{enumerate}
\item The uncertainty relations are violated by up to two orders of magnitude in thousands of experiments every single day.
\item Wavefunctions themselves are not real, but their components, mass and spin densities, are real.
\item Rotations in space generate complex numbers, which are not described in a Gibbs vector algebra.
\item Double slit interference experiments show two features: a discrete interaction spectrum with the slit system and a thermal broadening due to environmental conditions.
\item The fine structure constant and the electron rest mass describe the nuclear energy scale.
\item Closed shell nuclei are due to the geometrical arrangement of nuclear protons.
\end{enumerate}

On a personal note I think that fundamental Physics has entered a new stage of development, after the near inertia in the last thirty years. This is largely the merit of scientists working outside their core disciplines and motivated by nothing else but the curiosity, how things {\em really work}.
Finally, future developments in physics, based on this framework, could include the following elements:
\begin{itemize}
\item A non-relativistic theory of quantum electrodynamics making use of the constraint found for electromagnetic fields that the intensity {\em as well as the} frequency of the field must be linear with the energy of emission or adsorption.
\item A linear scaling many-electron theory for condensed matter making use of the result that many body effects can be encoded in a chiral optical potential.
\item A density functional theory of atomic nuclei using a high-density phase of electrons in the nuclear environment.
\end{itemize}

\section*{Acknowledgements}
Helpful discussions with Krisztian Palotas are gratefully acknowledged. The work was supported by the Royal Society London and the Canadian Institute for Advanced Research (CIFAR).

\end{document}